\documentstyle[prl,aps,twocolumn,epsf]{revtex}

\begin{document}

\draft

\title{Intermittent implosion and pattern formation of trapped
Bose-Einstein condensates with attractive interaction}

\author{Hiroki Saito and Masahito Ueda}
\address{Department of Physics, Tokyo Institute of Technology,
Tokyo 152-8551, Japan \\
and CREST, Japan Science and Technology Corporation (JST), Saitama
332-0012, Japan}

\date{\today}

\maketitle

\begin{abstract}
The collapsing dynamics of a trapped Bose-Einstein condensate (BEC) with
attractive interaction are revealed to exhibit two previously unknown
phenomena.
During the collapse, BEC undergoes a series of rapid implosions that
occur {\it intermittently} within a very small region.
When the sign of the interaction is suddenly switched from repulsive to
attractive, e.g., by the Feshbach resonance, density fluctuations grow to
form various patterns such as a shell structure.
\end{abstract}

\narrowtext

Bose-Einstein condensation (BEC) of trapped atomic vapor has been realized
in ${}^{87}{\rm Rb}$~\cite{Anderson}, ${}^{23}{\rm Na}$~\cite{Davis},     
${}^{1}{\rm H}$~\cite{Fried}, and ${}^{7}{\rm Li}$~\cite{Bradley}.        
The last species is unique in that it has a negative s-wave scattering
length, implying that the interactions between atoms are predominantly
attractive.
It has been believed that in a spatially uniform system, such atomic vapor 
would collapse into a denser phase.
However, when the system is spatially confined and when the number of BEC 
atoms is below a certain critical value $N_{\rm c}$, the zero-point motion
of the atoms serves as a kinetic obstacle against collapse, allowing a
metastable BEC to be
formed~\cite{Ruprecht,Fetter,Baym,Dalfovo96,Dodd,Ueda98,Kim,Wadati,UedaHuang,Huang,Saito}.
Just below $N_{\rm c}$, BEC may collapse via macroscopic quantum tunneling
~\cite{Kagan96,Shuryak,Stoof,Ueda98,Huepe,UedaHuang}, and above $N_c$, it
is predicted~\cite{UedaHuang} that the collapse will occur not globally,
but only locally near the center of BEC where the atomic density exceeds
a certain critical value.
Inelastic collisions also lead to the decay of BEC in regions where the
atomic density is very high~\cite{Dodd}.

In current experiments~\cite{Bradley}, there are abundant above-condensate
atoms that replenish the lost atoms, allowing BEC to grow again.
We may therefore expect collapse-and-growth cycles of BEC to occur.
Various mechanisms that give rise to these oscillations have been   
discussed~\cite{Sackett97,Sackett98,Kagan98,UedaHuang}.
And indeed, recent          
experiments~\cite{Sackett99} have suggested the occurrence of dynamic     
collapse-and-growth cycles of BEC, but the results have neither favored
nor excluded any one of these possible mechanisms.

The s-wave scattering length of atoms can be varied using the Feshbach
resonance~\cite{Inouye}, suggesting that not only the strength of the
interaction but also its sign can be controlled.
Using this technique, the sign of the interaction has recently been
successfully switched from positive to negative in BEC of ${}^{85}{\rm
Rb}$~\cite{JILA}.
Kagan {\it et al.}~\cite{Kagan97} have discussed a global collapse of the
condensate in such situations.

In this Letter, we predict two new phenomena associated with the collapse
of BEC.
One is {\it intermittent implosion}, in which the local collapses occur in
rapid sequence at the center of the condensate.
This phenomenon is caused by competition between the attraction of
atoms towards the trap center and the loss of atoms by inelastic
collisions.
While our analysis is based on the theory developed by Kagan {\it et
al.}~\cite{Kagan98}, this phenomenon is different from the
collapse-and-growth cycles and other fine structures predicted by them.
The other prediction is that of pattern formation in the atomic density,
following a sudden switch in sign of the interaction from repulsive to
attractive.
This phenomenon occurs because density fluctuations caused by the
change in sign of the interaction grow and self-focus due to the
attractive interactions.

We consider a system of Bose-condensed atoms with mass $m$ and s-wave  
scattering length $a$, confined in a parabolic potential.
The transition amplitude of the system from the initial state  
$\psi_{\rm i}$ to the final one $\psi_{\rm f}$ is expressed in
terms of path integrals as $\int_{\psi_i}^{\psi_f}  {\cal D} \psi {\cal D}
\psi^* e^{\frac{i}{\hbar}S[\psi, \psi^*]}$, where $S[\psi, \psi^*]$ is
given by
\begin{eqnarray} \label{S}
S[\psi, \psi^*] & = & N_0\hbar \int \!d{\bf r} \!\int\!dt 
\biggl[i\psi^*\frac{\partial}{\partial t} \psi 
+\frac{1}{2} \psi^* \nabla^2\psi \nonumber \\
& & - \frac{r^2}{2}\psi^*\psi -
\frac{g}{2} (\psi^* \psi)^2 \biggr].
\end{eqnarray}
Here the length, time, and $\psi$ are normalized in units of 
$d_0=(\hbar/m\omega_0)^{1/2}$, $\omega_0^{-1}$, and 
$(N_0/d_0^3)^{1/2}$, respectively, and $g\equiv4\pi N_0a/d_0$ is
the dimensionless strength of the interaction.
The wave function is then normalized to unity.
The most probable Feynman path that makes the action (\ref{S})
extremal satisfies the Gross-Pitaevskii (GP) equation~\cite{GP}
\begin{equation} \label{GP}
i \frac{\partial}{\partial t} \psi = -\frac{1}{2} \nabla^2 \psi +
\frac{r^2}{2} \psi + g |\psi|^2 \psi.
\end{equation}

The metastability of BEC with attractive interactions may be understood by
the Gaussian approximation~\cite{Fetter,Stoof,Ueda98}.
If we approximate the wave function as having a Gaussian form, the size of 
BEC --- $R$ --- obeys the equation of motion $\ddot R = -\partial V_{\rm
eff} / \partial R$, where the effective potential has the 
form $V_{\rm eff} = 3 (R^{-2} + R^2) / 2 + \gamma R^{-3} / 2$ with $\gamma
\equiv \frac{4N_0}{\sqrt{2\pi}} \frac{a}{d_0}$.
The effective potential $V_{\rm eff}$ has a local minimum when $|\gamma| <
\gamma_c \equiv 8 \cdot 5^{-5/4}$; therefore, the metastable BEC is
formed when there are less than the critical number of atoms $N_c
\equiv \frac{\sqrt{2\pi} d_0}{4 a} \gamma_c$.
We have performed a numerical integration of the GP equation (\ref{GP})
and have confirmed that the Gaussian approximation well describes
the dynamics of BEC.
However, the approximation breaks down when a rapid implosion takes
place.
Since we are interested in the behavior of the local implosion, we
numerically solve the GP equation without resorting to the Gaussian
approximation.

Because the peak density grows very high once the collapse begins, we
must include in the GP equation the atomic loss due to inelastic
collisions.
Following the treatment in Ref.~\cite{Kagan98}, we employ the GP equation
with loss processes as
\begin{equation} \label{GPloss}
i \frac{\partial}{\partial t} \psi = -\frac{1}{2} \nabla^2 \psi +
\frac{r^2}{2} \psi + g |\psi|^2 \psi - \frac{i}{2} \left( \frac{L_2}{2}
|\psi|^2 + \frac{L_3}{6} |\psi|^4 \right) \psi,
\end{equation}
where $L_2$ and $L_3$ denote the two-body dipolar and three-body
recombination loss-rate coefficients, respectively.
The two-body (three-body) loss-rate coefficients must be divided by two
(six) because of Bose statistics~\cite{factor}.
We assume that the atoms and molecules produced by inelastic collisions
escape from the trap without affecting the condensate.
Taking $\omega_0 = 2 \pi \times 144.5$ Hz\cite{Sackett99} and the
loss-rate coefficients from Refs.~\cite{Gerton,Moer}, we have $L_2 =
3.7 \times 10^{-7} N_0$ and $L_3 = 2.9 \times 10^{-10} N_0^2$.
To integrate the GP equation, we employ the finite difference method with
the Crank-Nicholson scheme~\cite{Ruprecht}.
Since the implosion is extremely rapid, we very carefully controlled the
stepsize to avoid error propagation during the implosion.

\begin{figure}[t]
\begin{center}
\leavevmode\epsfxsize=78mm \epsfbox{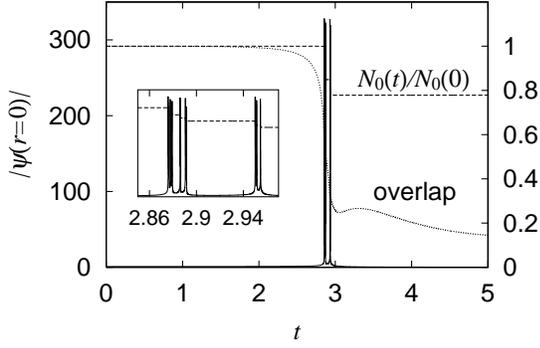}
\end{center}
\caption{
Time evolution of the wave function $|\psi(r = 0, t)|$ (solid curve
referring to the left axis), the number of atoms $N_0(t) / N_0(0)$ (dashed 
curve referring to the right axis), and the absolute squared overlap of
the wave function with the initial one $|\int d{\bf r} \psi^*({\bf r}, 0)
\psi({\bf r}, t)|^2$ (dotted curve referring to the right axis) with
two-body dipolar loss and three-body recombination loss.
We first prepared BEC in a metastable state slightly below the critical
point, and at $t = 0$, we increased $|a|$ so that $N_0 |a| / d_0$ exceeded
its critical value.
The inset enlarges the view of the intermittent implosion.
}
\label{f:implosion}
\end{figure}

Figure \ref{f:implosion} shows the time evolution of the peak height of
the wave function $|\psi(r = 0, t)|$ (solid curve), the number of BEC
atoms $N_0(t)$ (dashed curve), and the absolute squared overlap of the
wave function with the initial one $|\int d{\bf r} \psi^*({\bf r}, 0)
\psi({\bf r}, t)|^2$ (dotted curve) for $N_0(0) = 1260$, which is slightly
(0.7 \%) greater than $N_c$.
This gives an estimate of the ``condensate fraction'' that is measured by
the absorption or phase-contrast imaging (see discussions below).
We first prepared BEC in a metastable state that lay just below the
critical point, and then increased $|a|$ (or tightened the trap potential)
so that $N_0 |a| / d_0$ exceeded its critical value.
In the early stage of the collapse, the atomic density increases very
slowly and the inelastic collisions are unimportant.
At $t \simeq 2.87$ a rapid implosion breaks out, which is blown up in
the inset of Fig.~\ref{f:implosion}.
If the atomic loss were not included, the peak density would grow
unlimitedly, and the implosion would occur only once.
With the atomic loss, however, the implosion stops in a very short
time, and the peak density shows a pulse-like behavior.
This implosion occurs intermittently several times, and with each
implosion several tens of atoms are lost from the condensate.
As a result, the number of atoms decreases in a stepwise fashion,
eventually reaching approximately 78 \% of its initial value $N_0$.

Since the collisional loss of atoms takes place where the atomic density
is extremely high, the atomic loss primarily occurs within a very
localized spatial region $r \lesssim 0.01$.
The atomic loss is predominantly due to three-body recombination
rather than two-body dipolar decay, since the loss rate of the former
is proportional to the cube of the atomic density.
In fact, if the two-body loss is ignored, intermittent implosions occur
(data not shown).

The duration of the spike shown in Fig.~\ref{f:implosion} is typically
$\Delta t \sim 10^{-3}$.
The use of mean-field theory with such rapid dynamics can be justified
as follows.
The mean-field approximation is applicable for time scales longer than
$1 / g n$, which is of order 1 in the metastable state of BEC (Note that
the time is measured in units of $\omega_0^{-1}$).
In the region of implosion, the density $n$ becomes more than $10^4$ times 
as high as that of the metastable state, i.e., $1 / g n \lesssim 10^{-4}$,
and thus the mean-field theory is valid even with the rapid implosion if
the relevant time scale is longer than $10^{-4}$ --- which is the case for
the situation shown in Fig.~\ref{f:implosion}.
The gas parameter $n a^3$ is, on the other hand, $\sim 10^{-2}$ in the 
region of implosion, and the atoms are still acting in a
weakly interacting regime.

The pulse-like behavior of the peak atomic density may be interpreted
as follows.
Initially, the condensate has a negative pressure due to attractive
interactions and shrinks towards the central region.
When the peak density becomes $g |\psi|^2 \sim L_3
|\psi|^4 / 12$, i.e., $|\psi|^2 \sim 12g / L_3$, the collisional loss rate 
of the atoms becomes comparable to the accumulation rate of atoms at the
center.
Since the kinetic and interaction energies depend on the atomic
density and its square, respectively, the total energy increases upon the
loss of atoms~\cite{Kagan98}.
The atoms near the center of BEC thus acquire outward momentum, and the
pressure becomes positive.
The change in sign of the pressure may be qualitatively explained by
the Gaussian approximation.
Initially $-\partial V_{\rm eff} / \partial R \simeq 3 R^{-3} -
\frac{3}{2} |\gamma| R^{-4} < 0$, which corresponds to negative
pressure; later, however, the value becomes positive when the number of
atoms (or $|\gamma|$) decreases.
After the implosion, inward flow outside the region of the implosion
replenishes the peak density, turning the sign of the pressure again to
negative, which induces the subsequent implosion.

When the inward flow is insufficient to reverse the sign of the pressure
to negative, implosion ceases and the atoms are pushed outwards.
This phenomenon was predicted in Ref.~\cite{Kagan98}, and has recently
been observed at JILA~\cite{JILA} as an atom burst emanating from a
remnant condensate.
According to the estimation in the previous paragraph and our numerical
analysis~\cite{Saito}, the energy scale of implosion and subsequent
explosion is proportional to $g^2 / L_3$.
In the case of ${}^7{\rm Li}$, the mean energy of an ejected atom is
$\simeq 80$ $\mu{\rm K}$~\cite{Saito}.
The atoms and molecules are also scattered by three-body recombination, in
which the release energy is of the order 1 mK.
The experimental signature of the intermittent implosion should be a
series of bursts of atoms and molecules produced by the above two
mechanisms.

In the Rice experiments~\cite{Sackett99}, it has been observed that the
number of BEC atoms reduces to $10 \sim 20$ \% of the critical number
after the collapse, which is smaller than our theoretical evaluation
(dashed curve in Fig.~\ref{f:implosion}).
In Ref.~\cite{Sackett99}, however, only atoms around the peak of the
bimodal distribution are collected as the number of BEC atoms, while the
part of BEC atoms that expands broadly following the implosion is hidden
in the thermal cloud.
The number of BEC atoms that was measured experimentally is roughly
estimated from the absolute squared overlap of the wave function with the
initial metastable one (dotted curve in Fig.~\ref{f:implosion}) which
decreases to below 0.2.
Thus our result is consistent with the observation of
Ref.~\cite{Sackett99}.

The intermittent implosion should be distinguished from the two types of
oscillations discussed in Ref.~\cite{Kagan98}, i.e., the
collapse-and-growth cycles and the piecemeal collapses originating from
oscillations of the entire condensate.
The mechanism that causes the intermittent implosion is quite different
from that of those oscillations, as discussed above.
No intermittent implosion is discussed in Ref.~\cite{Kagan98}, as the
collisional loss rate used there is much larger than that used in this
Letter.

The scenario for the collapse of BEC in the presence of condensate growth
is, therefore, as follows.
The BEC grows by being fed by the above-condensate atoms, and when $N_0$
exceeds $N_c$, an implosion occurs that has the intermittent structure
shown in Fig.~\ref{f:implosion}.
Some of the atoms are lost in the implosion, but they are subsequently
replenished, giving rise to the collapse-and-growth cycles.
During these cycles, small collapses occur with period $\sim \omega_0^{-1}$
due to oscillations of the entire condensate.
These small collapses also have intermittent structures, which
we have confirmed by large-scale numerical simulations.

\begin{figure}[t]
\begin{center}
\leavevmode\epsfxsize=78mm \epsfbox{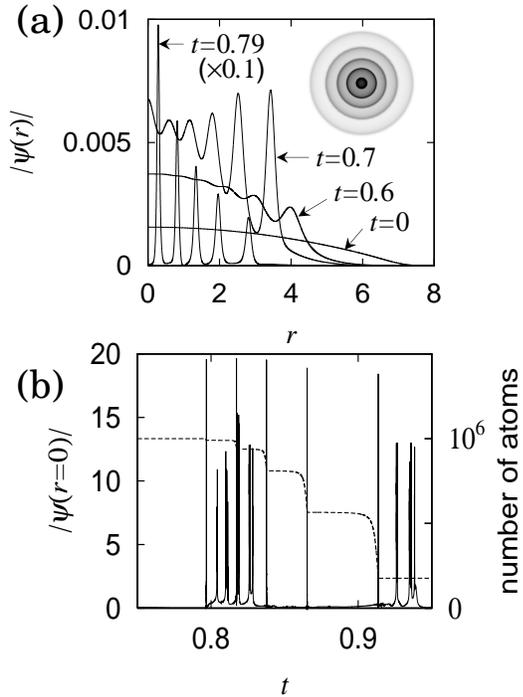}
\end{center}
\caption{
(a) Time evolution of the wave function $|\psi(r, t)|$.
A condensate of one million ${}^{23}{\rm Na}$ atoms was prepared, and at $t
= 0$, the s-wave scattering length was changed from $2.75$ nm to $-1$ nm.
The trap frequency is $\omega_0 = 100 \times 2 \pi$ ${\rm s}^{-1}$, and the 
loss-rate coefficients are described in the text.
The wave function at $t = 0.79$ is multiplied by 0.1.
The inset shows the gray-scale image of the column density at $t = 0.79$.
(b) Time evolution of the central height of the wave function $|\psi(r =
0, t)|$ (solid curve referring to the left axis) and the number
of atoms (dashed curve referring to the right axis).
}
\label{f:shell}
\end{figure}

We next consider the case in which BEC with repulsive interaction is
prepared, and the sign of the interaction is then suddenly switched to
attractive.
Such a situation has recently been realized at JILA~\cite{JILA} using
the Feshbach resonance.
Figure~\ref{f:shell} (a) shows the time evolution of the wave function,
where we assume that at $t = 0$ one million ${}^{23}{\rm Na}$ atoms having
an s-wave scattering length of $a = 2.75$ nm are condensed in the trap
with frequency $\omega_0 = 100 \times 2\pi$ ${\rm s}^{-1}$.
We then change the s-wave scattering length to a negative value of $-1$
nm.
We use the loss-rate coefficients provided in Ref.~\cite{Ketterle}, which
are $L_2 = 1.7 \times 10^{-8} N_0$ and $L_3 = 1.1 \times 10^{-10} N_0^2$.
When the interaction is switched to attractive, the atoms begin to
compress, as the initial wave function has been expanded due to the
repulsive interaction.
The inward flow gives rise to a ripple, which then grows up to be a series 
of pulses, as shown in Fig.~\ref{f:shell} (a) (the wave function at $t =
0.79$ is multiplied by 0.1).
The growth of the density fluctuations can be attributed to the attractive 
interaction, as the atoms tend to accumulate where the density is high.
The ripple is caused by the momentum acquired by the atoms sliding down
the trap potential.
In fact, it can be seen in Fig.~\ref{f:shell} (a) that the spaces between
adjacent pulses near the trap center are smaller than the outer spaces.
Since the potential energy becomes the kinetic energy as $\hbar^2 k^2 / 2
m \sim m \omega_0^2 R^2 / 2$, where $R$ is the initial dimension of the
wave function, the wavelength of the ripple is estimated to be $\lambda /
d_0 \sim 2 \pi d_0 / R$.

The inset in Fig.~\ref{f:shell} (a) shows a gray-scale image of the
column density integrated along the $z$ axis, $\int_{-\infty}^\infty
|\psi(x,y,z)|^2 dz$, at $t = 0.79$.
This quantity is proportional to the optical thickness in both absorption
and phase-contrast imaging, where the laser light propagates along the $z$
axis.
The concentric circles represent the formation of a shell structure in the 
atomic density.

The shell-structure formation is due to the instability of the initial
atomic distribution and the self-focusing effect, while the collisional
loss is unimportant for that formation since the atomic density is not so
high at the early stage of collapse when this phenomenon appears.
When implosion begins, however, the collisional loss of atoms begins to
play an important role.
Figure~\ref{f:shell} (b) shows the time evolution of the peak height of
the wave function $|\psi(r = 0, t)|$ (solid curve) and the number of BEC
atoms (dashed curve).
The peak density initially {\it decreases} because the atoms near the
center are attracted towards the innermost shell, as shown in
Fig.~\ref{f:shell} (a).
The shells move inward, and the first implosion occurs when the innermost
shell arrives at the center of the trap.
In Fig.~\ref{f:shell} (b), five implosions caused by the arrivals of the
shells at the center of the trap are shown.
The number of lost atoms associated with such implosions becomes larger
for the outer shell, since the number of atoms contained in each shell is
proportional to the square of its original radius.
The velocity of shells moving inward is roughly determined by the free
motion of atoms, and the collapse time is on the order of
$\omega_0^{-1}$.
Between these implosions, the smaller intermittent implosions discussed
above occur.
During each intermittent implosion, several tens of atoms are lost, a loss 
that cannot be discerned in Fig.~\ref{f:shell} (b) because the total
number of atoms is by far greater.

In the case of an axially symmetric trap, the pattern of the atomic
density arising from the change of the interaction is sensitive to the
asymmetry of the trap.
We performed numerical simulations and found that various patterns can be
formed.
For a pancake-shaped trap, the atomic motion associated with the change
in the interaction is larger in the axial direction than in the radial
direction.
As a result, the ripple arises in an axial direction, leading to a layered
structure.
For a cigar-shaped trap, the ripple arises in a radial direction, leading
to a cylindrical shell structure.
These structures undergo a complicated evolution and show various
patterns such as rings and clusters.
Intermittent implosions also occur for an axially symmetric trap.
The details of these phenomena will be reported elsewhere.

While we have analyzed the specific examples of ${}^{7}{\rm Li}$ and
${}^{23}{\rm Na}$,
the results should be valid for other atomic species in which the
s-wave scattering length can be varied, since the parameters $g$ and $L_3$
can be controlled by choosing $a$ and $N_0$.
With other values of $g$ and $L_3$, the behaviors may be qualitatively
different.
When $L_3$ is much larger than the value used here, no intermittent
implosion occurs.
Formation of a shell structure depends on the value of $g$ that is
proportional to $N_0$.
For instance, in the situation of Fig.~\ref{f:shell}, when the initial
number of atoms is $N_0 = 10^5$, the number of shells is reduced to two,
and when $N_0 = 10^4$, no shell structure appears.

In conclusion, we have predicted two new phenomena concerning the
collapsing dynamics of BEC with attractive interactions: intermittent
implosions and pattern formation in the atomic density.
The intermittent implosion occurs very rapidly compared with the time
scale of the trap frequency, and in a very localized region compared with
the characteristic size of the trap.
When the sign of the interaction is suddenly switched from repulsive to
attractive, the atomic density forms a shell structure in a spherically
symmetric trap, and various patterns are formed for an axially symmetric
trap.

This work was supported by a Grant-in-Aid for Scientific Research (Grant
No. 11216204) by the Ministry of Education, Science, Sports, and Culture
of Japan, and by the Toray Science Foundation.

\newpage

\end{document}